\documentclass[twocolumn,showpacs,showkeys]{revtex4}

\usepackage{amsmath}
\usepackage{amssymb,amsfonts}
\usepackage{bm}
\usepackage[mathcal]{euscript}
\usepackage{graphicx}
\usepackage{psfrag}
\usepackage{subfigure}

\newcommand{\comment}[1]{}

\newcommand{\ie}{\textit{i.e.}}

\newcommand{\mathnotation}[2]{\newcommand{#1}{\ensuremath{#2}}}

\newcommand{\transp}[1]{{#1}^T}			
\newcommand{\Order}[1]{\ensuremath{\mathrm{O}\!\l(#1\r)}}

\newcommand{\goodgap}{%
	\hspace{\subfigtopskip}%
	\hspace{\subfigbottomskip}}

\renewcommand{\l}{\left}			
\renewcommand{\r}{\right}			
\mathnotation{\pd}{\partial}			
\mathnotation{\ee}{{\mathrm e}}			
\mathnotation{\imi}{{\mathrm{i}}}		
\mathnotation{\ldef}{\mathrel{\raisebox{.069ex}{:}\!\!=}}
\mathnotation{\rdef}{\mathrel{=\!\!\raisebox{.069ex}{:}}}
\mathnotation{\dint}{\,{\mathrm{d}}}		

\mathnotation{\grad}{\nabla}			
\renewcommand{\div}{\grad\cdot}			
\mathnotation{\curl}{\grad\times}		
\mathnotation{\lapl}{\nabla^2}			

\renewcommand{\time}{t}				
\mathnotation{\xc}{x}				
\mathnotation{\xv}{{\bm{\xc}}}			
\mathnotation{\velc}{v}				
\mathnotation{\velv}{{\bm{\velc}}}		
\mathnotation{\flow}{\Phi}			
\mathnotation{\Mc}{M}				
\mathnotation{\Mt}{\mathbb{\Mc}}		
\mathnotation{\sdim}{n}				

\mathnotation{\advsc}{\phi}			
\mathnotation{\dens}{\rho}			
\mathnotation{\Diffc}{D}			
\mathnotation{\Difft}{\mathbb{\Diffc}}		
\mathnotation{\diff}{\kappa}			
\mathnotation{\tdiff}{\tilde{\diff}}		
\mathnotation{\Pe}{\mathrm{Pe}}			

\mathnotation{\edir}{e}				
\mathnotation{\edirv}{\mathbf{\edir}}		
\mathnotation{\ediru}{\hat{\edir}}		
\mathnotation{\ediruv}{\hat{\mathbf{\edir}}}	
\mathnotation{\ediruinf}{\ediru^\infty}		
\mathnotation{\ediruvinf}{\ediruv^\infty}	
\mathnotation{\edirt}{\tilde{\edir}}
\mathnotation{\edirtv}{\tilde{\mathbf{\edirv}}}
\mathnotation{\sdir}{{\mathrm{s}}}		
\mathnotation{\sdirv}{\mathbf{s}}		
\mathnotation{\sdiru}{\hat{\sdir}}		
\mathnotation{\sdiruv}{\hat{\sdirv}}		
\mathnotation{\sdiruinf}{\sdiru^\infty}		
\mathnotation{\sdiruvinf}{\sdiruv^\infty}	
\mathnotation{\sdirt}{\tilde{\sdir}}
\mathnotation{\sdirtv}{\tilde{\sdirv}}
\mathnotation{\udir}{{\mathrm{u}}}		
\mathnotation{\udirv}{\mathbf{u}}		
\mathnotation{\udiru}{\hat{\udir}}		
\mathnotation{\udiruv}{\hat{\udirv}}		
\mathnotation{\udiruinf}{\udiru^\infty}		
\mathnotation{\udiruvinf}{\udiruv^\infty}	
\mathnotation{\udirt}{\tilde{\udir}}
\mathnotation{\udirtv}{\tilde{\udirv}}
\mathnotation{\mdir}{{\mathrm{m}}}		
\mathnotation{\mdirv}{\mathbf{m}}		
\mathnotation{\mdiru}{\hat{\mdir}}		
\mathnotation{\mdiruv}{\hat{\mdirv}}		
\mathnotation{\mdiruinf}{\mdiru^\infty}		
\mathnotation{\mdiruvinf}{\mdiruv^\infty}	
\mathnotation{\mdirt}{\tilde{\mdir}}
\mathnotation{\mdirtv}{\tilde{\mdirv}}

\mathnotation{\lac}{a}				
\mathnotation{\lav}{{\bm{\lac}}}		
\mathnotation{\lavs}{\lav_\sdir}		
\mathnotation{\lavz}{\lav_0}			
\mathnotation{\timelac}{\time_0}		
\mathnotation{\metric}{g}			
\mathnotation{\detmetric}{|\metric|}		
\mathnotation{\lyapexp}{\lambda}
\mathnotation{\nugr}{\Lambda}			
\mathnotation{\nugrpara}{\widetilde\nugr}	%
\mathnotation{\anugr}{\mathcal{K}}		
\mathnotation{\upar}{u}				
\mathnotation{\mpar}{m}				
\mathnotation{\uarclen}{\hat{\upar}}		
\mathnotation{\spar}{s}				
\mathnotation{\sarclen}{\hat{\spar}}		
\mathnotation{\ks}{k}				

\mathnotation{\gradlac}{\nabla_0}
\mathnotation{\divlac}{\grad_0\cdot}
\DeclareMathOperator{\cdivlac}{div_0}
\mathnotation{\curllac}{\grad_0\times}
\mathnotation{\lapllac}{\nabla_0^2}
\mathnotation{\advsclac}{\advsc_0}
\mathnotation{\denslac}{\dens_0}
\mathnotation{\tdenslac}{\tilde{\dens}}		
\mathnotation{\tadvsclac}{\tilde\advsc}
\mathnotation{\Diffclac}{\Diffc_0}
\mathnotation{\Difftlac}{\Difft_0}
\mathnotation{\metriclac}{g_0}			
\mathnotation{\detmetriclac}{|\metriclac|}
\mathnotation{\detmetriclacpara}{\widetilde{\metric}_0}

\mathnotation{\dunusfunc}{c}
\mathnotation{\pdu}{\pd_\upar}
\mathnotation{\pdm}{\pd_\mpar}

\newcommand{\adeq}{advection--diffusion equation}
\newcommand{\Peclet}{P\'{e}clet}
\newcommand{\twoD}{2D}
\newcommand{\threeD}{3D}
\newcommand{\sline}{$\sdiruv$-line}

\begin{document}

\title{Advection--Diffusion in Lagrangian Coordinates}
\author{Jean-Luc Thiffeault}
\email{jeanluc@mailaps.org}
\affiliation{Department of Applied Physics and Applied Mathematics, Columbia
  University, New York, NY 10027}
\altaffiliation[Present address: ]{Department of Mathematics, Imperial College
  London, SW7 2AZ.}

\begin{abstract}
The \adeq\ can be approximated by a one-dimensional diffusion equation in
Lagrangian coordinates along the directions of compression of fluid elements
(the stable manifold).  This result holds in any number of dimensions, for a
velocity field with chaotic trajectories, with an error proportional to the
square root of the diffusivity.  After some time, the one-dimensional equation
becomes invalid, but by that time a large fraction of the scalar variance has
decayed.
\end{abstract}

\maketitle

\section{Introduction}
\label{sec:intro}

The importance of transport and mixing processes in the physical sciences
cannot be overestimated.  Applications abound in the geophysical sciences such
as oceanography, atmospheric science, and geology, as well as in astrophysics,
chemistry, and general fluid dynamics.  Because it spans so many disciplines,
a comprehensive understanding of the detailed manner in which mixing proceeds
is of paramount importance.

In most physically relevant applications it is the case that the mixing
(diffusion) process depends on the stirring (advection) phase to create large
gradients that then allow diffusion to act.  Without some form of enhanced
stirring, diffusion is powerless in the face of weak gradients.  The
enhancement may be provided by turbulence in the flow, but that is not
necessary: for even simple laminar flows, fluid trajectories can display
sensitivity to initial conditions (chaos) that stretches fluid elements
exponentially---a phenomenon termed \emph{chaotic advection}~\cite{Aref1984}.
This stretching produces the necessary large gradients of the advected
quantity, leading to chaotic mixing.  Our focus will be on such smooth,
chaotic flows.

There are two main results in this paper.  First, we show that the \adeq\ in
Lagrangian coordinates can be reduced to a one-dimensional diffusion equation
along the stable manifold of the flow (the direction along which fluid
elements are compressed).  The diffusion coefficient of the reduced equation
grows exponentially in time, reflecting the enhancement to diffusion due to
chaotic trajectories; the reduced equation itself is an approximation valid to
exponential accuracy in time.  The enhancement to diffusion is not uniform
along the stable manifold: the amount of stretching of fluid elements is
anticorrelated with the curvature of the stable
manifold~\cite{Tang1996,Thiffeault2001}.  The second result is that this
approximate, one-dimensional, equation breaks down after a certain time
because of a buildup of large gradients along the neglected direction of
stretching.  By that time, a significant fraction of the scalar variance has
decayed, so the one-dimensional equation is often sufficient for describing
physical systems.

\section{The Reduced Advection--Diffusion Equation}
\label{sec:derivation1d}

In the present section we give an account of the manner in which the \adeq\
expressed in Lagrangian coordinates can be collapsed to a one-dimensional
diffusion equation.  The result holds regardless of dimension (two, three, or
more as can arise in kinetic equations), and the approximations involved are
accurate to order~$\Pe^{-1/2}$, where the \Peclet\ number is~$\Pe\ldef
L\,\velc/\Diffc$, with~$L$ and~$\velc$ characteristic spatial and velocity
scales, and~$\Diffc$ the diffusivity; in many common problems,~$\Pe>10^{8}$.
Though it has been realized for some time that the enhanced diffusion proceeds
along the stable (contracting) direction in both
smooth~\cite{Tang1996,Antonsen1996} and
turbulent~\cite{Chertkov1998,Son1999,Balkovsky1999} flows, here we make a
series of approximations and use a geometrical
constraint~\cite{Tang1996,Thiffeault2001,Thiffeault2002} to rewrite the \adeq\
as an actual one-dimensional diffusion equation along the stable manifold.
This is more than mere formalism, because it allows the identification of the
role of invariant structures of the flow in the mixing process
(Section~\ref{sec:phys1d}).  Solving the one-dimensional equation also allows
us to show that it must eventually break down (Section~\ref{sec:breakdown}).

The \adeq\ for a compressible velocity field~$\velv(\xv,\time)$ on
an~$\sdim$-dimensional bounded domain is
\begin{equation}
	\frac{\pd\advsc}{\pd\time} + \velv\cdot\grad\advsc
	= \frac{1}{\dens}\,\div\l(\dens\,\Difft\cdot\grad\advsc\r),
	\label{eq:advdiffE}
\end{equation}
where the scalar field~$\advsc(\xv,\time)$ is the concentration of the
quantity advected, $\Difft(\xv,\time)$ is the diffusivity tensor, and~$\dens$
is the density of the fluid (which satisfies the continuity equation).  The
advected scalar~$\advsc$ can represent for example temperature, salinity, or
the concentration of a solute such as helium.  The velocity field~$\velv$ can
be obtained by solving the Navier--Stokes equation, or it can be a model
system designed to mimic the qualitative features of a realistic flow.  The
trajectory of a fluid element is given by the
solution~$\xv(\time,\timelac;\lav)$ to
\begin{equation}
	\frac{\pd\xv}{\pd\time} = \velv(\xv,\time),
	\quad
	\xv(\time=\timelac,\timelac;\lav) = \lav,
\end{equation}
so that~$\xv(\time,\timelac;\lav)$ is the coordinate transformation at
time~$\time$ from the comoving Lagrangian ($\lav$) to stationary Eulerian
($\xv$) coordinates.  For smooth velocity fields, the transformation is smooth
and invertible.

By construction, in Lagrangian coordinates the advection term drops out, and
the \adeq~\eqref{eq:advdiffE} takes the form
\begin{equation}
	{\Bigl.\frac{\pd\advsclac}{\pd\time}\Bigr|}_{\lav}
	= \frac{1}{\denslac}\,\divlac\l(\denslac\,
		\Difftlac\cdot\gradlac\advsclac\r),
	\label{eq:advdiffL}
\end{equation}
where~$0$ subscripts denote Lagrangian
quantities: $\advsclac(\lav,\time)=\advsc(\xv(\time,\timelac;\lav),\time)$,
$\denslac(\lav) = |\Mt|\,\dens(\xv(\time,\timelac;\lav),\time)$,
\hbox{$\gradlac = \pd/\pd\lav$}, and
\begin{equation}
	\Difftlac = \Mt^{-1}\cdot\Difft\cdot(\Mt^{-1})\transp{{}},
	\label{eq:Diffclac}
\end{equation}
where the tangent map (or Jacobian)~$\Mt$ has
components~\hbox{${\Mc^i}_q=\pd\xc^i/\pd\lac^q$}.  The~$\lav$ subscript on the
time derivative in~\eqref{eq:advdiffL} is a reminder that the derivative is
taken with~$\lav$ held fixed, as opposed to the derivative
in~\eqref{eq:advdiffE} which has~$\xv$ fixed.

The covariance of~\eqref{eq:advdiffE} and~\eqref{eq:advdiffL} (\ie, the form
of the right-hand side of both equations is the same) suggests the definition
of a metric \comment{Don't forget the~$\timelac$.}
\begin{equation}
	\metric \ldef \diff\,\Difft^{-1}
	\label{eq:metric}
\end{equation}
on the tangent space at~$\xv$, where the dimensional
function~$\diff(\lav,\time)$ is chosen such that~$\detmetric=1$.  (For
constant~$\Difft$, $\diff$ is also constant.)  The metric~$\metric$ is a
proper Riemannian metric, since it is symmetric and positive-definite; it
induces a metric~$\metriclac$ on the tangent space at~$\lav$,
\begin{equation}
	\metriclac = \transp{\Mt}\cdot\metric\cdot\Mt
	= \diff\,\Difftlac^{-1}.
	\label{eq:metriclac}
\end{equation}
Because it is symmetric and positive-definite, the metric~${\metriclac}$ can
be written in terms of its real-positive eigenvalues~$\nugr_\sigma^2$ and
orthonormal eigenvectors~$\ediruv_\sigma$ as
\begin{equation}
	{\metriclac} = \sum_{\sigma=1}^{\sdim}
		\nugr_\sigma^2\,\ediruv_\sigma
		\,\ediruv_\sigma\,.
	\label{eq:metricdiag}
\end{equation}
The~$\nugr_\sigma(\time,\timelac;\lav)$ are called \emph{coefficients of
expansion}; without loss of generality, they are ordered such
that~\hbox{$\nugr_1 \ge \nugr_2 \ge \ldots \ge \nugr_\sdim$}.  For a very wide
class of flows, including incompressible flows,~$\nugr_\sdim$ (the smallest
coefficient of expansion) is associated with an exponentially contracting
direction, so that~\hbox{$\nugr_\sdim\ll 1$}.  The characteristic
directions~$\ediruv_\sigma(\time,\timelac;\lav)$ converge exponentially to
their time-asymptotic
values~$\ediruvinf_\sigma(\timelac;\lav)$~\cite{Goldhirsch1987,%
Thiffeault2002}.  The direction of fastest contraction ($\sigma=\sdim$) plays
a distinguished role in our development, so we emphasize its importance by
denoting it by the letter~`$\sdir$' (for ``stable'').  Thus,~$\ediruv_\sdim$
and~$\nugr_\sdim$ are written~$\sdiruv$ and~$\nugr_\sdir$.  Similarly, we will
later use the letters~`$\udir$' (for ``unstable'') for the direction of
fastest stretching ($\sigma=1$); for three-dimensional systems, we
use~`$\mdir$' (for ``median'') when referring to direction~$\sigma=2$, which
may be stretching, contracting, or neutral.

Replacing~$\Difftlac$ by~$\diff\metriclac^{-1}$ using~\eqref{eq:metriclac},
and substituting the diagonal form~\eqref{eq:metricdiag} of the metric
into~\eqref{eq:advdiffL}, we find
\begin{equation}
	{\Bigl.\frac{\pd\advsclac}{\pd\time}\Bigr|}_{\lav}
	= \sum_{\sigma=1}^\sdim
		\frac{1}{\denslac}\,\divlac\l(\denslac\,\diff\,
		\nugr_\sigma^{-2}\,\ediruv_\sigma
		\,\ediruv_\sigma\cdot\gradlac\advsclac\r).
	\label{eq:advdiffL3}
\end{equation}
Since~\hbox{$\diff \sim \Pe^{-1}\ll 1$}, for short times the right-hand side
of~\eqref{eq:advdiffL3} can be neglected completely.  However, in a chaotic
flow, at least some of the~$\nugr_\sigma$ achieve exponential behavior after a
moderate time and can overcome the small diffusivity.  Because the inverse
of~$\nugr_\sigma$ enters~\eqref{eq:advdiffL3}, the direction of fastest
contraction~$\nugr_\sdir=\nugr_\sdim$ dominates, and eventually we
have~\hbox{$\Pe^{-1}\nugr_\sdir^{-2} \sim \Order{1}$}.  Assuming
\hbox{$\nugr_\sdir \sim \exp(\lyapexp_\sdir\,\time)$} this occurs roughly at
time
\begin{equation}
  \time_1-\timelac \simeq \frac{1}{2|\lyapexp_\sdir|}\,\log\Pe,
  \label{eq:t1def}
\end{equation}
after which we can approximate~\eqref{eq:advdiffL3} by
\begin{equation}
	{\Bigl.\frac{\pd\advsclac}{\pd\time}\Bigr|}_{\lav}
	= \frac{1}{\denslac}\,\divlac\l(\denslac\,\diff\,
		\nugr_\sdir^{-2}\,\sdiruv\,\sdiruv\cdot\gradlac\advsclac\r)
	+ \mathrm{O}(\nugr_{\sdim-1}^{-2}).
	\label{eq:advdiffL4}
\end{equation}
The relative error we are making if we neglect
the~$\mathrm{O}(\nugr_{\sdim-1}^{-2})$ terms is of
order~${(\nugr_\sdir/\nugr_{\sdim-1})}^2$.
Since~$\Pe$ is assumed very large, by the time~\hbox{$\nugr_\sdir^{-2} \sim
\Pe$} the coefficient of expansion~$\nugr_\sdir$ is well into the exponential
regime, and we have~$\nugr_\sdir\ll\nugr_{\sdim-1}$, if the two coefficients
are exponentially separated.  Thus, for the approximation~\eqref{eq:advdiffL4}
to hold, it is sufficient to require that the two smallest Lyapunov exponents
differ, which we assume to be the case.  In a~\threeD\ autonomous flow, we
have~$\nugr_{\sdim-1}=\nugr_\mdir\sim 1$ because of the presence of a zero
Lyapunov exponent~\cite{Eckmann1985}.  Hence, the relative error
in~\eqref{eq:advdiffL4} is of order~\hbox{$\nugr_\sdir^2 \sim \Pe^{-1} \ll
1$}.  For a \twoD\ time-dependent flow, the error is even smaller, being of
order~\hbox{${(\nugr_\sdir/\nugr_\udir)}^2 \sim (\Pe^{-1}/\nugr_\udir^2) \ll
\Pe^{-1}$}.  We conclude that in most physically relevant applications the
neglect of the higher-order terms in~\eqref{eq:advdiffL4} is correct to a
very high degree of accuracy.  However, in Section~\ref{sec:breakdown} we will
show that this neglect is not justified for long times.

Physically, the leading-order term in~\eqref{eq:advdiffL4} reflects the
enhanced diffusion due to the presence of a dominant contracting direction:
along that direction huge gradients are created by the stretching and folding
action of the flow, leading to an exponential increase in the efficiency of
diffusion.  For all practical purposes the other directions can be ignored, as
long as the smallest Lyapunov exponents are nondegenerate, as assumed.

Since the eigenvector~$\sdiruv$ converges to its asymptotic value~$\sdiruvinf$
at a rate~$(\nugr_\sdir/\nugr_{\sdim-1})$~\cite{Goldhirsch1987,%
Thiffeault2002}, we may replace~$\sdiruv(\time,\timelac;\lav)$
in~\eqref{eq:advdiffL4} by~$\sdiruvinf(\timelac;\lav)$.  The error committed
is proportional to~$\nugrpara_\sdir\sim\Pe^{-1/2}$, larger than the
order~$\Pe^{-1}$ error in~\eqref{eq:advdiffL4}.  The characteristic
direction~$\sdiruvinf(\timelac;\lav)$ can be integrated to yield the
\emph{stable manifold}~$\lavs(\sarclen)$ (or~\sline) \comment{I'm leaving
out the~$\timelac$ dependence in~$\lavs$} through a point~$\lavz$,
\begin{equation}
	\frac{\pd\lavs(\sarclen)}{\pd\sarclen}
		= \sdiruvinf(\timelac;\lavs(\sarclen)),
	\qquad \lavs(0)=\lavz,
	\label{eq:stablemandef}
\end{equation}
where~$\sarclen$ denotes the arc length along the stable manifold.  The stable
manifold is central to our development, and we will
reformulate~\eqref{eq:advdiffL4} so that it represents a diffusion equation
along this manifold.

To achieve this goal two more approximations are needed.  It was shown in
Ref.~\cite{Thiffeault2002} that~\hbox{$\sdiruv\cdot\gradlac\log\nugr_\sdir$}
converges exponentially to a time-independent function, at a
rate~$\max(\nugr_\sdir\,,{\nugr_\sdir}/{\nugr_{\sdim-1}})$.  This convergence
implies that, along an~\sline\ given by~$\lavs(\sarclen)$, we can choose
an arbitrary reference point~$\lavs(0)=\lavz$ and express the
time-dependence of~$\nugr_\sdir(\time,\timelac;\lavs(\sarclen))$ as
\begin{equation}
	\nugr_\sdir(\time,\timelac;\lav) =
		\nugrpara_\sdir(\timelac;\lav)\,
		\nugr_\sdir(\time,\timelac;\lavz),
	\label{eq:nugrslinedecomp}
\end{equation}
where~$\lav$ and~$\lavz$ are on the same~\sline\ and~$\nugrpara_\sdir$ is a
time-independent function.  We will discuss the physical significance of the
function~$\nugrpara_\sdir$, which we call the parallel coefficient of
expansion, in Section~\ref{sec:phys1d}.  Neglecting the time-dependence of
$\nugrpara_\sdir$ does not increase the overall error.

We now rewrite~\eqref{eq:advdiffL4} as
\begin{equation}
	{\Bigl.\frac{\pd\advsclac}{\pd\time}\Bigr|}_{\lav}
	\simeq \frac{1}{\denslac}\,\divlac\l(\denslac\,\tdiff\,
		\,\sdirtv\,\sdirtv\cdot\gradlac\advsclac\r),
	\label{eq:advdiffL5}
\end{equation}
where we have defined
\begin{equation}
	\tdiff \ldef
		\diff\,\nugr_\sdir^{-2}(\time,\timelac;\lavz),
	\qquad
	\sdirtv \ldef \nugrpara_\sdir^{-1}\,\sdiruvinf\,.
	\label{eq:tdiffsdirtvdef}
\end{equation}
In Ref.~\cite{Thiffeault2002}, it was demonstrated that for a chaotic flow in
any number of dimensions,~$\sdirtv$ must satisfy the constraint
\begin{equation}
	\cdivlac\sdirtv = \nugrpara_\sdir^{-1}
	\bigl(\cdivlac\sdiruvinf
		- \sdiruvinf\cdot\gradlac\log\nugrpara_\sdir\bigr)
		= 0,
	\label{eq:constraint}
\end{equation}
where
\begin{equation}
	\cdivlac\sdirtv \ldef \detmetriclac^{-1/2}\,\divlac
		\bigl(\detmetriclac^{1/2}\,\sdirtv\bigr)
	\label{eq:covdivlac}
\end{equation}
is the covariant divergence of~$\sdirtv$~\cite[p.~49]{Wald}.  Upon using the
constraint~\eqref{eq:constraint} in~\eqref{eq:advdiffL5}, we obtain
\begin{equation}
	{\Bigl.\frac{\pd\advsclac}{\pd\time}\Bigr|}_{\lav}
	\simeq \frac{1}{\tdenslac}\,\sdirtv\cdot\gradlac
		\l(\tdenslac\,\tdiff\,\,\sdirtv\cdot
		\gradlac\advsclac\r),
	\label{eq:advdiffL6}
\end{equation}
where
\begin{equation}
	\tdenslac \ldef {\detmetriclacpara}\!{}^{-1/2}\,\denslac\,,
\end{equation}
and~$\detmetriclacpara(\lav)$ is defined analogously to~$\nugrpara_\sdir$
in~\eqref{eq:nugrslinedecomp},
\begin{equation}
	|\metriclac(\lav,\time)| =
		\detmetriclacpara(\lav)\,|\metriclac(\lavz,\time)|.
	\label{eq:detmetricslinedecomp}
\end{equation}
The decomposition~\eqref{eq:detmetricslinedecomp} is possible because the
derivative of the metric determinant along the~\sline\ converges exponentially
to a time-independent value~\cite{Thiffeault2002}.  The
factor~$\detmetriclacpara$ in~$\tdenslac$ arises from the metric determinant
in the covariant divergence~\eqref{eq:covdivlac}, and is absent in
incompressible flows.  Thus, after all the approximations have been made, the
overall error is proportional
to~$\max(\nugr_\sdir,(\nugr_\sdir/\nugr_{\sdim-1}))$, which scales
as~$\Pe^{-1/2}$.

Finally, we define a derivative along~$\sdirtv$,
\begin{equation}
	\frac{\pd}{\pd\spar} \ldef \sdirtv\cdot\gradlac\,,
	\label{eq:pdspar}
\end{equation}
where~$\spar$ is a new parameter along the~\sline\ that differs from the arc
length~$\sarclen$ because of the weight~$\nugrpara_\sdir^{-1}$
in~\eqref{eq:tdiffsdirtvdef}.  Equation~\eqref{eq:advdiffL6} then achieves
the simple form
\begin{equation}
	{\Bigl.\frac{\pd\advsclac}{\pd\time}\Bigr|}_{\lav}
	\simeq \frac{1}{\tdenslac}\,
		\frac{\pd}{\pd\spar}\l(\tdenslac\,\tdiff\,
		\,\frac{\pd\advsclac}{\pd\spar}\r).
	\label{eq:advdiff-1d}
\end{equation}
Equation~\eqref{eq:advdiff-1d} is a one-dimensional diffusion equation along a
line defined by~$\lavs(\spar)$.  The effective
diffusivity~$\tdiff(\lavs(\spar),\time)$ diverges exponentially in time,
so that even for a microscopic physical diffusivity~$\diff$ the gradients
along~$\sdirtv$ are eventually smoothed.  Note that it is only in terms of the
parameter~$\spar$ that we obtain a one-dimensional diffusion equation; in
terms of any other parameter (other than a uniform rescaling of~$\spar$),
equation~\eqref{eq:advdiff-1d} contains a ``drift'' term proportional
to~$\divlac\sdiruvinf$.

\section{Physical Interpretation}
\label{sec:phys1d}

We now address the physical meaning of the various constituents of the
one-dimensional diffusion equation~\eqref{eq:advdiff-1d}.  The essence of the
simplification in Section~\ref{sec:derivation1d} lies in expressing the \adeq\
in Lagrangian coordinates in terms of a single parameter along a stable
manifold: with an appropriate choice of parameter, namely~$\spar$, the
equation collapses to a simple diffusion equation.

There are three main ingredients involved in the one-dimensional
equation~\eqref{eq:advdiff-1d}: (i) The exponential part of the coefficient of
expansion,~$\nugr_\sdir(\time,\timelac;\lavz)$; (ii) the stable
manifold,~$\lavs(\sarclen)$; and (iii) The parallel coefficient of
expansion,~$\nugrpara_\sdir$.  We discuss each in turn.

\textbf{(i)} The exponential part of the coefficient of
expansion,~$\nugr_\sdir(\time,\timelac;\lavz)$, is responsible for the
enhanced diffusion coefficient~$\tdiff$.  The variation of the
function~$\nugr_\sdir$ across~\sline{}s along unstable directions is very
steep (see Section~\ref{sec:breakdown}) and is the cause of the poor
convergence of Lyapunov exponents~\cite{Tang1996,Thiffeault2002}.  Thus, the
diffusion rates on two neighboring~\sline{}s can be quite different in
magnitude, though similar in character
because~$\nugr_\sdir(\time,\timelac;\lavz)$ is independent of~$\spar$ and so
gives only a time-dependent overall scaling of the diffusion coefficient
in~\eqref{eq:advdiff-1d}.

\textbf{(ii)} The~\sline\ is straightforward to
compute~\cite{Tang1996,Thiffeault2001}; it has a fine structure and densely
fills an ergodic region.  If we treat the stable manifold as a material line
to be advected by the flow, then the elements on that line will converge
exponentially together as~$\time\rightarrow\infty$.  Thus, if these initial
elements carry different concentrations of~$\advsc$, the gradients of~$\advsc$
will grow exponentially along these trajectories.  It is thus natural for the
stable manifold to play a central role in the chaotic mixing process.  In
fact, it is the capacity of the stable manifold to fill an entire chaotic
region that allows for a thorough mixing~\cite{Pierrehumbert2000}.

When evolved \emph{backward} in time, a small ``test parcel'' (of a scale
comparable to the length at which dissipation acts) will be stretched and
folded and will give a filament of fluid that samples the initial
data~$\advsclac$~\cite{Pierrehumbert2000}.  This filament will tend to trace
out the stable manifold.  This is the Lagrangian analogue of the phenomenon of
``asymptotic directionality'' invoked to discuss the behavior of a material
line as it is evolved forward in time by a
flow~\cite{Alvarez1998,Giona1998,Giona1998b,Muzzio2000}.

\textbf{(iii)} The parallel coefficient of expansion~$\nugrpara_\sdir$, also
has deep implications for chaotic mixing.  The time-dependent part
of~$\nugr_\sdir$, whose role was explored in~(i) above, is uniform on
an~\sline.  In contrast, the parallel coefficient of
expansion~$\nugrpara_\sdir$ is constant in time (though it depends
on~$\timelac$ for unsteady flows) but varies along the stable manifold.  Since
the full coefficient of expansion~$\nugr_\sdir$ gives the rate of contraction
of an initial infinitesimal ellipsoid, we must regard~$\nugrpara_\sdir$ as a
\emph{deviation} from the reference rate of contraction given
by~$\nugr_\sdir(\time,\timelac;\lavz)$.  Thus, regions of anomalously
large~$\nugrpara_\sdir$ are associated with fluid elements that contract
slower than other elements on the same~\sline.

The size of~$\nugrpara_\sdir$ has an impact on diffusion, because the enhanced
diffusion is due to the contraction of fluid elements.  The effect is embodied
in~$\pd/\pd\spar$: because of~\eqref{eq:tdiffsdirtvdef} and~\eqref{eq:pdspar},
we have~\hbox{${\pd}/{\pd\spar} = \nugrpara_\sdir^{-1}\,{\pd}/{\pd\sarclen}$},
where~$\sarclen$ is the arc length along~$\sdiruvinf$.  Hence, in regions of
large~$\nugrpara_\sdir$, gradients of~$\advsclac$ are \emph{smaller} when
expressed in the coordinate~$\spar$; the reverse is true for regions of
small~$\nugrpara_\sdir$.  A large value of~$\nugrpara_\sdir$ is thus a
hindrance to diffusion.

\begin{figure*}
\centering
\subfigure[]{
	\psfrag{th}{{$\sarclen$}}
	\psfrag{log L}{{$\log\nugrpara_\sdir$}}
	\includegraphics[width=3.17in]{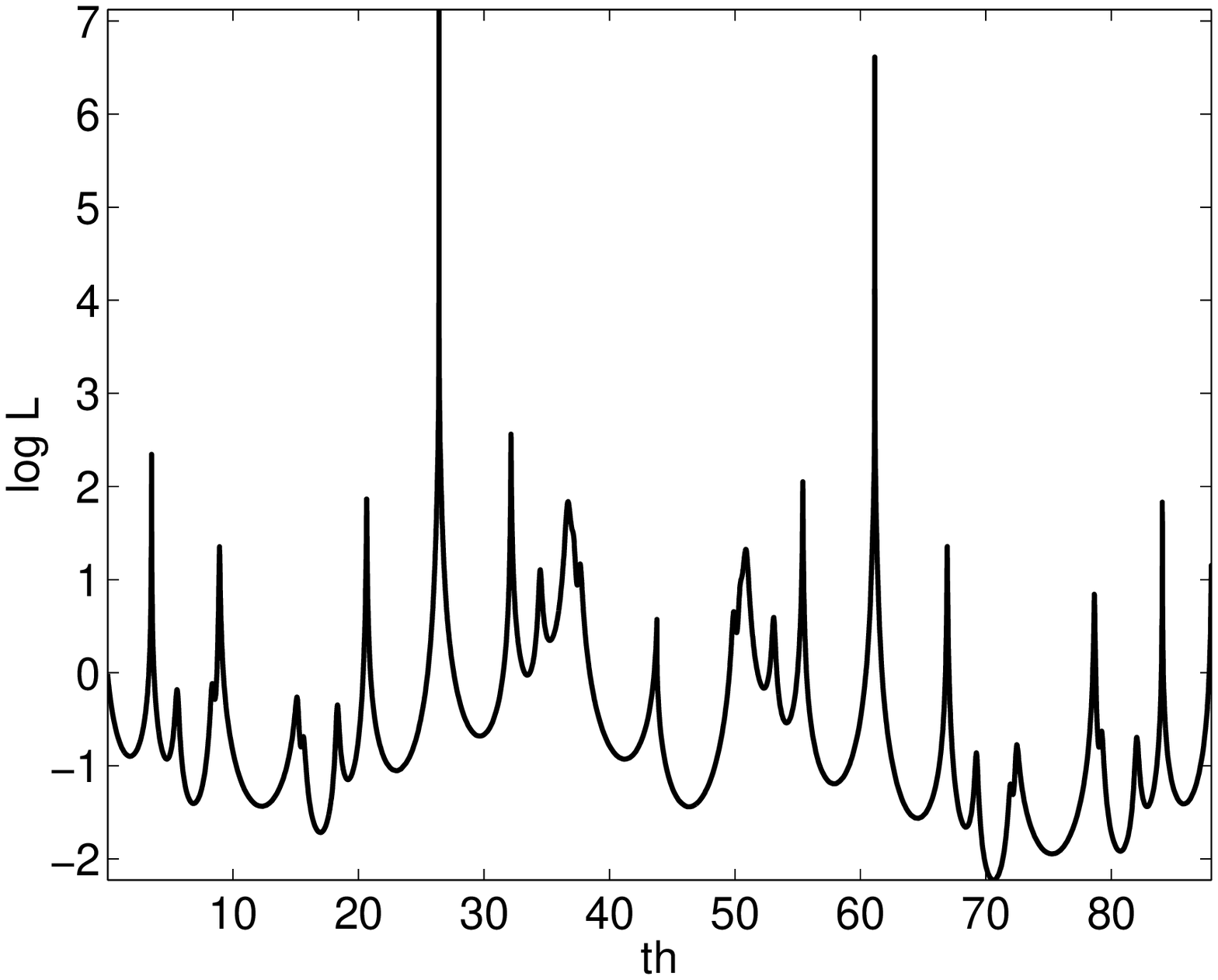}
	\label{fig:Lamt_vs_tauh}
}\goodgap
\subfigure[]{
	\psfrag{N}{$N$}
	\psfrag{L2}{{\small $\nugrpara^{-2}_\sdir$}}
	\psfrag{log L}{{$\log\nugrpara_\sdir$}}
	\includegraphics[width=3.25in]{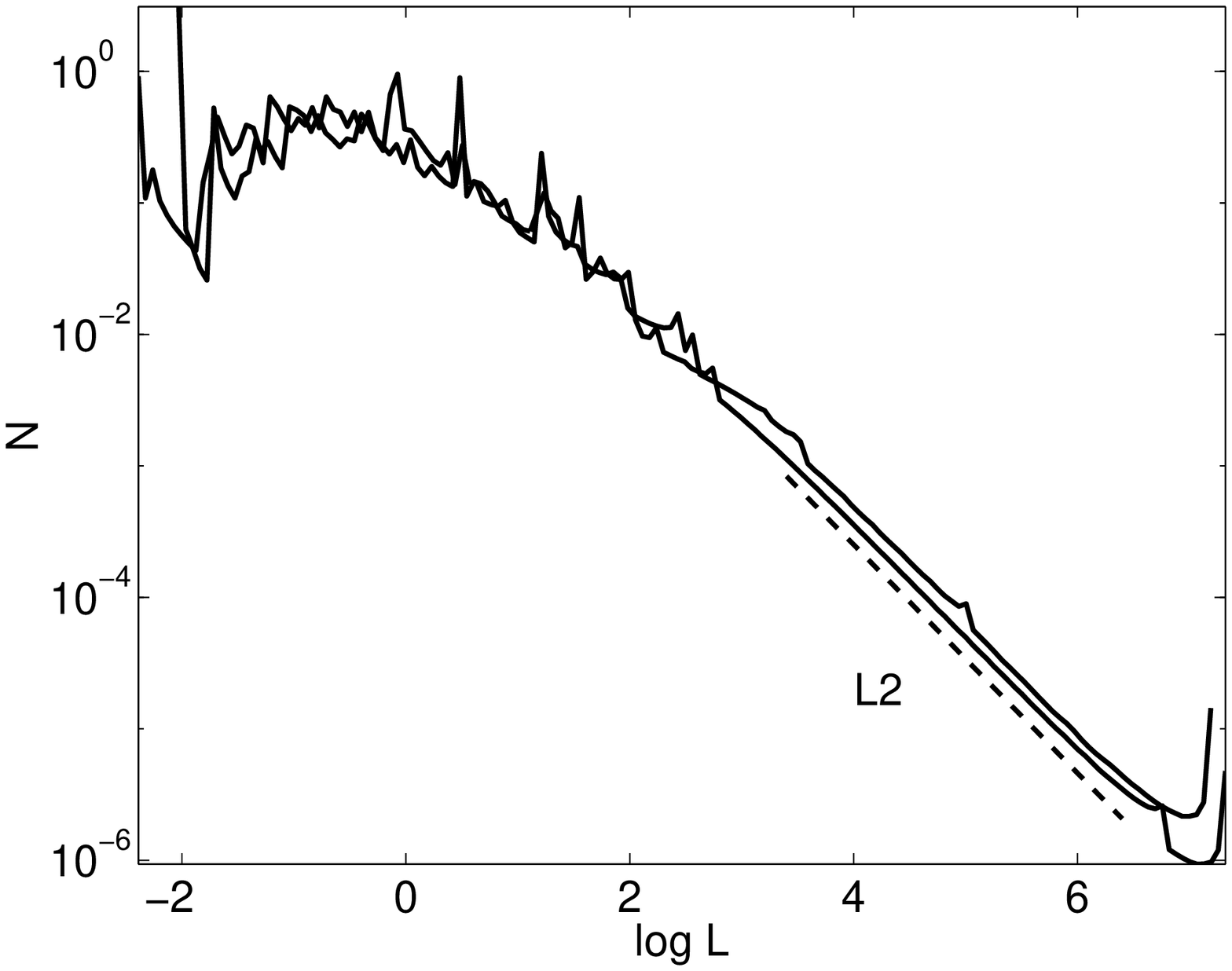}
	\label{fig:pdfLamt}
}
\caption{(a) $\log\nugrpara_\sdir$ as a function of arc length~$\sarclen$
along a typical~\sline.  (b) Normalized PDF of~$\nugrpara_\sdir$ on
two~\sline{}s of equal length.  For comparison, the dashed line
denotes~$\nugrpara_\sdir^{-2}$.  The PDFs on the two~\sline{}s appear very
similar.}
\end{figure*}

But what range of values does~$\nugrpara_\sdir$ achieve?  In
Fig.~\ref{fig:Lamt_vs_tauh} we plot~$\log\nugrpara_\sdir$ against the arc
length~$\sarclen$ on an~\sline.  The cellular flow of Solomon and
Gollub~\cite{Solomon1988,Wiggins} is used in the computation (with parameter
values~$A=B=\omega=k=1$); it is meant to mimic a chain of oscillating
convection rolls.
The parallel coefficient of expansion~$\nugrpara_\sdir$ is seen to vary over
five orders of magnitude.  Thus, there exist many exceptional fluid elements
that have very slow rates of contraction.

The important factor, then, is how frequently these large values
of~$\nugrpara_\sdir$ occur along the~\sline.  Figure~\ref{fig:pdfLamt} shows
the probability distribution function (PDF) of~$\log\nugrpara_\sdir$ along
two~\sline{}s: the tail of the distribution is clearly an exponential,
decaying as~$\exp(-2\log\nugrpara_\sdir)=\nugrpara_\sdir^{-2}$.  This is
surprising because for this flow the distribution of the finite-time Lyapunov
exponents themselves has a Gaussian tail, as opposed to a ``fat''
(exponential) tail.  In fact, we have observed the~$\nugrpara_\sdir^{-2}$
exponential law for the tail of the distribution in several flows, even flows
such as the random wave flow~\cite{Antonsen1996} (also known as the sine flow)
that have no KAM regions.  Thus, the anomalously large values
of~$\nugrpara_\sdir$ cannot only be due to sticking to KAM surfaces, though
that is certainly a factor.  Rather, there are inherent paths in the flow that
fluid elements can follow to avoid stretching, and every ``final'' fluid
element contains a mixture of such paths.  \comment{What is the distribution
of~$\nugr_\sdir(\time,\timelac;\lavz)$?}  That the tails of the PDF
of~$\log\nugrpara_\sdir$ fall off exponentially rather than as a Gaussian
suggests that these exceptional events are important to the long-time decay of
scalar variance, in the same manner that the initial alignment of the scalar
gradient along the~\sline\ matters.  The statistics of these regions of low
stretching is related to the distribution of curvature of material
lines~\cite{Schekochihin2002,Thiffeault2003strcurv}.

\section{Breakdown of the Approximation}
\label{sec:breakdown}

We now show that the neglect of the higher-order terms in~\eqref{eq:advdiffL4}
is not justified for long times, a fact we believe has not been appreciated
before.  To do that, we compare the magnitude of the terms that were neglected
in going from~\eqref{eq:advdiffL3} to~\eqref{eq:advdiffL4} and show that they
become important after a certain time.  To simplify the argument, we assume
the flow is incompressible, and that the diffusivity~$\diff$ is constant in
space and time, so that~$\tdiff$ in~\eqref{eq:tdiffsdirtvdef} is constant
along the \sline\ and depends only on which particular stable manifold we are
dealing with [see~(i) in Section~\ref{sec:phys1d}].  We label the dependence
on the particular manifold under consideration by a parameter~$\upar$, meant
to represent displacements along an unstable (stretching) direction~$\udiruv$
of the flow.  For an initial condition~\hbox{$\advsclac(\spar,0;\upar) =
\exp(\imi\ks\spar)$} (we set the initial time~$\timelac$ to zero), the
one-dimensional diffusion equation~\eqref{eq:advdiff-1d} has solution
\begin{equation}
  \advsclac(\spar,\time;\upar) = \exp\l(\imi\ks \spar
  - \ks^2\,\time\,\anugr(\time;\upar)\r),
  \label{eq:superexpsol}
\end{equation}
where
\begin{equation}
  \anugr(\time;\upar) \ldef \frac{1}{\time}
  \int_{0}^\time\tdiff(\time';\upar)\dint\time'
  \label{eq:anugrdef}
\end{equation}
is the time-average of~$\tdiff(\time;\upar)$.  From
Eq.~\eqref{eq:tdiffsdirtvdef},~$\tdiff$ is proportional
to~\hbox{$\nugr_\sdir^{-2} \sim \exp(2|\lyapexp_\sdir|\time)$},
so~\eqref{eq:superexpsol} gives a \emph{superexponential} decay
of~$\advsclac$.

Consider now the variation of~$\advsclac(\spar,\time;\upar)$ along the unstable
direction~$\udiruv$,
\begin{equation}
  \pdu\advsclac = -\ks^2\,\time\,\pdu\anugr(\time;\upar)\,\advsclac,
  \label{eq:pduadvsc}
\end{equation}
where~$\pdu\ldef\udiruv\cdot\gradlac$.  The second derivative
of~$\advsclac(\spar,\time;\upar)$ is
\begin{equation}
  \pdu^2\advsclac = \l[\l(\ks^2\,\time\,\pdu\anugr(\time;\upar)\r)^2
    - \ks^2\,\time\,\pdu^2\anugr(\time;\upar)\r]\advsclac.
  \label{eq:pdu2advsc}
\end{equation}
Keeping only the first term in~\eqref{eq:pdu2advsc} (the second term does not
dominate), we take the ratio of the~$\sigma=1$ (unstable direction) term to
the~$\sigma=3$ (stable direction) term in~\eqref{eq:advdiffL3},
\begin{equation}
  \varepsilon \ldef 
  \frac{\bigl\lvert\diff\,\nugr_\udir^{-2}\l(\ks^2\,\time\,\pdu\anugr\r)^2
    \advsclac\bigr\rvert}
       {\bigl\lvert\diff\,\nugr_\sdir^{-2}\,\ks^2\advsclac\bigr\rvert}
  = {\bigl\lvert\nugr_\udir^{-1}\nugr_\sdir\,\ks\,\time\l(\pdu\anugr\r)
    \bigr\rvert^2}\,;
  \label{eq:ratio}
\end{equation}
The ratio~$\varepsilon$ must be small to justify the approximation leading
from~\eqref{eq:advdiffL3} to~\eqref{eq:advdiffL4}.  Because~$\tdiff$ grows
exponentially, we have~\hbox{$\anugr \sim
\diff\,\nugr_\sdir^{-2}/2\lvert\lyapexp_\sdir\rvert\time$}, so
that
\begin{equation}
  \pdu\anugr
  \sim \frac{\diff}{2\lvert\lyapexp_\sdir\rvert\time}\,\pdu\nugr_\sdir^{-2}
  = \frac{\diff}{\lvert\lyapexp_\sdir\rvert\time}\,\nugr_\sdir^{-2}\,
  \pdu\log\nugr_\sdir^{-1}.
  \label{eq:pduK}
\end{equation}
Since~$\nugr_\sdir$ is in the exponential regime, we have
that~\cite{Thiffeault2002}
\begin{equation}
  \pdu\log\nugr_\sdir
  = \dunusfunc\,\nugr_\udir\,,
  \label{eq:pdunugrs}
\end{equation}
where~$\dunusfunc$ is a function of space and has only algebraic behavior in
time: it neither grows nor decays exponentially in time.  The exponential
growth in~\eqref{eq:pdunugrs} is a reflection of the exponential separation of
fluid trajectories in the chaotic flow.  Equation~\eqref{eq:pdunugrs} applies
for almost all initial conditions in essentially all smooth flows and maps (it
is sufficient that the velocity field be
thrice-differentiable~\cite{Thiffeault2002}), except in highly exceptional
cases such as uniformly stretching systems (e.g., Arnold's cat map, or flow
around a hyperbolic point), for which~$\varepsilon\equiv0$; the
superexponential solution~\eqref{eq:superexpsol} then persists forever.

After inserting~\eqref{eq:pduK}--\eqref{eq:pdunugrs} in~\eqref{eq:ratio},
we obtain finally
\begin{equation}
  \varepsilon \sim \l\lvert\ks\dunusfunc\,\lyapexp_\sdir^{-1}\,
  \diff\,\nugr_\sdir^{-1}\r\rvert^2
  \sim \l\lvert\Pe^{-1}\,\nugr_\sdir^{-1}\r\rvert^2.
    \label{eq:epsasym}
\end{equation}
Assuming that~$\nugr_\sdir^{-1} \sim \exp(|\lyapexp_\sdir|\time)$, we conclude
that~$\varepsilon$ becomes~$\Order{1}$ at time
\begin{equation}
  \time_2 \simeq
  \frac{1}{\lvert\lyapexp_\sdir\rvert}\,\log\Pe = 2\time_{1}\,,
  \label{eq:t2def}
\end{equation}
where~$\time_1$ is the time where diffusion sets in, defined
by~\eqref{eq:t1def}.

In three dimensions, we must take into account the~\hbox{$\sigma=2$} (the
`\mdir' or `median' direction) term in~\eqref{eq:advdiffL3}.  Its role depends
on the type of direction it represents.  If it is a stretching or neutral
direction, then the estimate presented above holds, and diffusion along
the~$\sigma=2$ direction becomes important at roughly the same time~$\time_2$
given by~\eqref{eq:t2def}.  If~\hbox{$\sigma=2$} is a contracting direction,
we must modify the asymptotic behavior~\eqref{eq:pdunugrs}
to~\hbox{$\pdm\log\nugr_\sdir = \bar\dunusfunc\,(\nugr_\sdir/\nugr_\mdir)$},
as described in~\cite{Thiffeault2002}.  This leads to a time~\hbox{$\time_2'
\ldef (1/2\lvert\lyapexp_\mdir\rvert)\log\Pe$}, after which the~$\sigma=2$
term in~\eqref{eq:advdiffL3} can no longer be neglected.  Depending on the
magnitude of~$|\lyapexp_\mdir|$, $\time_2'$ may come before or
after~$\time_2$.  The breakdown of~\eqref{eq:advdiff-1d} thus occurs
at~$\min(\time_2,\time_2')$, which we will simply call~$\time_2$.  In two
dimensions, we ignore~$\time_2'$ because there is no median direction.

The overall picture is thus as follows.  For time~\hbox{$\time < \time_1$},
$\advsclac$ is approximatively constant because the exponential growth of
gradients has not yet overcome the small diffusivity.  For~\hbox{$\time_1 <
\time < \time_2$}, Eq.~\eqref{eq:advdiff-1d} applies and the scalar variance
decays superexpontially.  For~\hbox{$\time>\time_2$}, the neglect of
the~\hbox{$\sigma=1$} (and~$\sigma=2$ in~\threeD) terms
in~\eqref{eq:advdiffL3} is no longer justified, and the one-dimensional
diffusion equation~\eqref{eq:advdiff-1d} must be abandoned: to characterize
the advection--diffusion process, the full Eq.~\eqref{eq:advdiffL3} must be
solved.

The reason that the diffusion along~$\udiruv$ cannot be neglected
after~$\time_2$ is that the tiny, exponentially-decaying effective diffusivity
associated with~$\udiruv$ allows the creation of extremely large gradients
of~$\advsclac$ along that direction.  These gradients are a consequence of the
superexponential solution~\eqref{eq:superexpsol}.  Eventually, the gradients
overcome the exponentially-decaying effective diffusivity.  In essence, the
superexponential solution contains the seeds of its own destruction: the
superexponentiality comes from the factor~$\anugr$ in~\eqref{eq:superexpsol},
but that factor exhibits extremely rapid variation along the stretching
direction, with no diffusivity to oppose it (until~$\time\simeq\time_2$).

The breakdown of the one-dimensional equation helps to understand a
discrepancy between the superexponential decay predicted by that equation and
the observed exponential decay of the scalar variance in typical chaotic and
turbulent flows.  In Ref.~\cite{Antonsen1996} a superexponential decay was
derived for individual fluid trajectories, and an exponential decay was
recovered by averaging over the initial angle between the scalar gradient and
the stable direction.  \citet{Fereday2002} pointed out that the decay of the
variance in the (nonuniform) baker's map is exponential for long times, even
though in their case the angle between the scalar gradient and the stable
direction is always zero, so that no averaging can be done.  Our demonstration
of the termination of the superexponential regime after some time shows that
there is hope of restoring the more appropriate, exponential decay of the
variance, but we have yet to show this because a technique of solution for the
Eq.~\eqref{eq:advdiffL3} must be found.  For the Kraichnan--Kazantsev model of
homogeneous, isotropic turbulence (delta-correlated in time), the decay of
variance is known to be exponential~\cite{Chertkov1998,Son1999,Balkovsky1999},
though the predicted rate is too rapid when compared with nonhomogeneous
systems~\cite{Fereday2002} which can have slowly-decaying eigenfunctions.

Even though it ceases to be valid after some time, the one-dimensional
equation is still relevant, because during the time~\hbox{$\time_1 < \time <
\time_2$} a very large fraction of the variance can get dissipated, owing to
the very fast decay of the superexponential.

\begin{acknowledgments}
The author thanks S. Childress, A. H. Boozer and D. Lazanja for helpful
discussions.  This work was supported by the National Science Foundation and
the Department of Energy under a Partnership in Basic Plasma Science grant,
No.~DE-FG02-97ER54441.
\end{acknowledgments}


\end{document}